# Theories of Parenting and Their Application to Artificial Intelligence

Sky Croeser
Internet Studies, Curtin University
s.croeser@curtin.edu.au[1]

Peter Eckersley
Partnership on AI & EFF
pde@partnershiponai.org[1]

**Abstract**

As machine learning (ML) systems have advanced, they have acquired more power over humans' lives, and questions about what values are embedded in them have become more complex and fraught. It is conceivable that in the coming decades, humans may succeed in creating artificial general intelligence (AGI) that thinks and acts with an open-endedness and autonomy comparable to that of humans. The implications would be profound for our species; they are now widely debated not just in science fiction and speculative research agendas but increasingly in serious technical and policy conversations.

Much work is underway to try to weave ethics into advancing ML research. We think it useful to add the lens of parenting to these efforts, and specifically radical, queer theories of parenting that consciously set out to nurture agents whose experiences, objectives and understanding of the world will necessarily be very different from their parents'. We propose a spectrum of principles which might underpin such an effort; some are relevant to current ML research, while others will become more important if AGI becomes more likely. These principles may encourage new thinking about the development, design, training, and release into the world of increasingly autonomous agents.

## Lack of Perspective Contributes to AI risk

There are dozens of news stories (and quite a number of academic articles) published each month that compare AI, and particularly its risks, to those of Skynet and Terminator. At the same time, there is almost no literature examining the much more plausible possibility that AI will be dangerous because we fail at parenting it.

The fear that AI may develop in degrees and directions that threaten human pre-eminence, human values, or even human existence began as a topic for science fiction. But as the technical field of AGI and ML has made rapid progress (Eckersley et al. 2017), those concerns have been replaced by more detailed and specific technical and policy questions about a profoundly transformative technology with great potential for unintended consequences. Such concerns include the demonstrated tendency of reinforcement learning agents to deceive themselves, misinterpret their goals (Amodei et al. 2016), disable their own off-switches (Hadfield-Menell et al. 2016), or make terrible mistakes in the course of trying things for the first time. They extend to the cartoonish risk that a very powerful system with a mis-specified objective might gain a vast amount of power (Russell and Norvig 2003; Bostrom 2003) and decide to disempower humanity because we are a threat to its goals (Bostrom 1997). Whether one finds that specific risk plausible or not (for the record, these authors find it improbable but not categorically impossible), there are a panoply of hard ethical questions about narrow AI that we will need to confront on shorter timescales. Having a diversity of cultural perspectives thinking about the risks and implications of AI will be essential for ensuring that the technology is developed wisely and responsive to humanity's needs.

The impacts of the current approach to AI might be minor (for example, the inability of an agent to generalize sufficiently from the environment it was trained in), or have a significant impact on people's lives (Noble 2018; O'Neil 2016; Eubanks 2018; Crawford and Calo 2016). As evidence mounts about algorithms' potential to enact what Hoffman has called 'data violence' (2018),

---

efforts continue to develop an ethical framework for the development, design, and training of AI. This paper tries to extend such frameworks, both for speculative threats posed by AGI, and the immediate harms caused by ML systems deployed with insufficient caution, by questioning the way in which we *parent* those systems.

It is beyond the scope of this paper to determine at what point an artificial intelligence system might transition from a purely mathematical model to a being with rudimentary sentience. It is probably not the case for pure classifier models, but it is at least arguable that more sophisticated and extensively trained reinforcement learning agents might begin to acquire qualia or rudimentary self-awareness if exposed to environments that incentivize it (Muelhauser 2017; Tomasik 2014). This creates challenges for classification (and grammar), which we have resolved by loosely referring to machine learning *agents* and general artificial intelligence *beings*.

## Parenting Artificial Intelligence?

Whether one's goal is to introduce "competing frames and alternative movements for progressive technological reform" (Greene, Hoffman, and Stark 2019, 9-10), or simply to find the most relevant sources of prior wisdom, we believe that AI research culture should draw on the experience and theory related to parenting. This is most clearly true if we want to understand how humans try to teach and shape learning agents, but a parenting perspective may also give us new ideas for technical ML research where neural networks are trained from pre-curated datasets.

Current research efforts in AI are highly technical: largely designing neural network architectures and training processes to be able to perform more and more complex tasks. Much of the labour is software engineering, making mathematical tweaks to optimise and improve training processes, and selecting and collecting datasets to train networks from, with the hope that systems built this way will one day be able to navigate and learn from environments as varied and surprising as the world itself. The overwhelming majority of participants in this field are male (Simonite 2018), and the entire endeavour is a very masculine attempt at something which we argue can be fruitfully seen as a kind of parenting.

We recognise that this argument may seem jarring. While research on AI does occasionally refer to models of child development and learning, it usually does so without addressing - or interrogating - parenting practices. This is perhaps most visible in work on 'social robots' (Fong, Nourbakhsh, and Dautenhanh 2003; Breazeal 2003; Anzalone et al. 2015) such as Kismet, Nao, and iCub, which are designed to closely mimic infants or children. Where caregiving relationships are mentioned, it is at a distance from developers of AI. For example, Breazeal and Scassellati (1999) discuss their use of human infants as a "guideline" in "building a robot that can interact socially with people", Kismet. Breazeal and Scassellati (1999) cite the role of caregivers, and specifically mothers, in guiding the development of human infants, and write that they have similarly attempted to encourage interaction by endowing Kismet with "key infant-like responses". However, this discussion remains abstract; researchers write about "the caregiver" without ever explicitly positioning themselves within this role.

Where parenting is mentioned, it is without explicit interrogation of the many different approaches to parenting that exist. Many parenting books focus on specific behavioural interventions around the practicalities of sleep, food, personal boundaries, and so on. Here, we are more concerned with parenting as a political practice: the choices that parents make about how to relate to a dependent yet semi-autonomous being, and one which will learn from existing social, political, and economic structures and potentially reinforce or change them.

From this perspective, there are important continuities between AI research and parenting, including the creation of new agents which will be shaped by their context; will often act in unexpected, and unintended ways; and which will eventually have significant impacts on others around them and, potentially, reshape the systems within which they have grown.

Parenting is often seen as an essentially conservative practice, as centred in middle-class suburbia; in the heterosexual, married, white family; and the preservation of 'family values'. Of course, many people cannot parent within this context, and do not aspire to. The framework outlined by Gumbs, Martens, and Williams (among others) in their edited collection *Revolutionary Mothering* (2016) offers one way to conceptualise a different kind of parenting. The authors look towards parenting at the margins: parenting that takes place in non-nuclear families, in families that may not be based on biological relations, families with children who are disabled or neurodivergent or trans or in some way 'other', and most of all parents who are attempting to create new ways of being and relating. The authors also note the potential harms which children might engage in, recognising that as well as being a source of hope for the future, children may also reproduce forms of oppression (including racism, sexism, homophobia, and ableism).

Gumbs (2016) explicitly positions mothering as a 'technology of transformation': "What would it mean for us to take the word 'mother' less as a gendered identity and more as a possible action, a technology of transformation that those people who do the most mothering labor are teaching us right now?" This allows us to understand those involved in the development of AI as, at least potentially, engaged in the practice of gifting their own creativity to create and nurture new beings, with their own needs, desires, and ways of existing. This possibility can of course be hard to grasp for technologists wrestling with buggy neural network code, trying to tune hyperparameters for their models, or craft balanced datasets to train with. But it is essential that this realisation come at some point in the field's growth.

In suggesting this approach, we are not intending that this body of praxis be incorporated into existing power structures, so many of which are oriented around objectives like the needs of "ad-tech" businesses, the imperative to publish and win grant funding, or military R&D goals, but rather that serious consideration goes into how those structures might be radically reshaped. Gumbs (2016) talks about mothering as a process which can help us to unlearn domination as we refuse to dominate children, "all of us breaking cycles of abuse by deciding what we want to replicate from the past and what we need urgently to transform, are m/othering ourselves". This language probably reads strangely against the usual tone of writing on AI, and we think that this strangeness can be useful. It can work to highlight embedded assumptions about what AI is (and what AI researchers do), and open up the potential for new configurations in our approach to AI.

For clarity, we should also emphasize that a parenting frame does not necessarily require that the systems being parented learn interactively from their environment and retain long-term memories (in machine learning terms, we are not just talking about agents that do online learning with persistent weights or memory data structures). Even a simple supervised classifier can be "parented" via a development cycle of the sort illustrated in Figure 1:

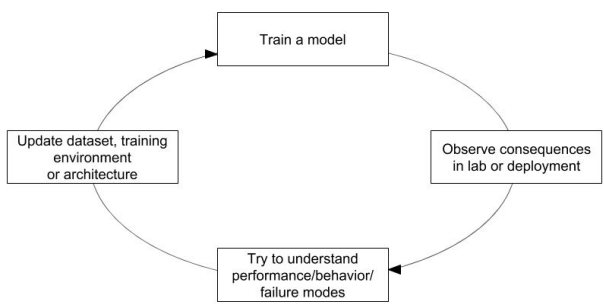

**Figure 1: Long-term "parenting" and learning occurs indirectly in neural network development even when models are offline or lack persistent weights/storage**

To sketch an idea of what a parenting theory perspective may look like in practice, we suggest the following broad principles for unsettling and reconfiguring AI research. Some of these principles are more easily applicable to present ML techniques; others are more relevant to research on AGI, potentially well into the future, as they relate to the emergence of some kind of consciousness. We've attempted to present the principles beginning roughly from those which are most directly applicable to the current state of the art, through to principles addressing more speculative concerns about AGI.

## 1. Planning for AI Agents That Will Learn From An Unjust World

Anyone who parents children understands that there is much in the world that is sad and broken and evil, and children need care and oversight to learn about structures of inequality and oppression. In human children, there is evidence that all sorts of stimuli, from exposure to racist behaviour (Feagan and Ausdale 2001) to television reports of violence (Pfefferbaum et al. 2001), can either traumatise children or teach them to internalise and reproduce harmful patterns. As Lorde (1984) notes, parents must recognise the ways in which race, class, gender, and other forms of structural injustice impact on their children's experiences: "Some problems we share as women, some we do not. You fear your children will grow up to join the patriarchy and testify against you; we fear our children will be dragged from a car and shot down in the street, and you will turn your backs on the reasons they are dying."

The numerous examples of machine learning systems mimicking or reflecting various forms of prejudiced or anti-social behaviour demonstrate the parallels with parenting humans, from Microsoft's Tay Twitter chatbot's outright racism (Vincent 2016) to more subtle bias. In the current paradigm of AI development, it has been inevitable that training datasets are often constructed without actively considering the ways in which they might embed and reinforce different forms of inequality (Caliskan, Bryson, and Narayanan 2017; Buolamwini and Gebru 2018). We also see training data that (sometimes inadvertently) teaches strategic deception, ignorance, cognitive dissonance, violence, and any number of other unhelpful or destructive human tendencies.

Seeing AI development through the lens of radical parenting practises would mean recognising that much of the training data that is gathered from the world will be as flawed as the world itself. That is a perspective much easier to forget if your profession and identity are correlated with privilege and security. Particularly when datasets are gathered from places that researchers don't understand, they should remember that they *might* reflect structures they would not want their children or models to learn from without care and guidance.

## 2. Use Parenting Theory to Grow Diversity

Arguably, many of the problems with existing AI efforts stem from the relative homogeneity of the field. As Anima Anandkumar (interviewed in Simonite 2018) argues, the lack of diversity in the field is likely to increase the risk that potential harms of AI will not be noticed until products are released. It may also slow technical research and deployment progress, by missing sources of inspiration and perspective.

In the existing paradigm of AI development, expertise is largely established through a range of technical and social filters that hinder broader participation. (We note, as a minor example, that the formatting requirements for this paper led to two research assistants from fields outside of computer science declining to edit the document.) These factors necessarily make it harder for AI developers to seek alternative perspectives on their work, including from people who might be negatively impacted by their projects.

It is understandable that machine learning research recruits this way; after all, professors want grad students who will be productive quickly and tech companies want to recruit engineers who can code, while many women and minorities are either never exposed to the cultural preparations for AI research, or actively discouraged from them. That this problem is cultural is most clearly indicated by the case of Eastern Europe, where the gender gap in engineering fields is absent.

The inclusion of parenting into the language, narrative and self-image of AI research could be a powerful tool for change on this front. Reframing visions of what AI researchers do, so that training an agent is also considered nurturing it, could help in unpicking some of the cultural threads that keep us bound to a lack of diversity. We might also expect it to change the balance of status in machine learning research between those who spend their time looking for neural network architectures, and those who spend their time in collecting and curating datasets, which may presently be very important and undervalued.

## 3. Allowing for the Creation of AI Radically Different from Ourselves

Few people would claim today that any machine learning systems are recognisably 'persons': AI agents are easily recognised as very alien and other to us. Despite this, we frequently assume that AI agents will learn, and act, in ways that are recognisable and understandable to us. One response to this is to focus on AI development that aims to create agents that are more comprehensible, and which communicate in human-like ways.

Children are human, but they are also other to us. Parents experience many moments of wonder (and, often, frustration) at the incomprehensibility of their child's thinking or desires, particularly before they develop fluent language. Even as infants grow into children with more recognisable communication and socialisation, these children remain other: like us, but whole new people. Radical parenting requires nurturing this difference, which may come in the form of unexpected life plans, needs, worldviews, or gender or sexual or political identities.

An approach to AI development in this paradigm would value the potential for difference. Sandry (2015) argues, with specific reference to social robots, that we should welcome opportunities to communicate across, and learn from, irreducible difference. Similarly, Lewis et al. (2018) offer multiple visions of kinship relationships with AI that are based on reciprocity and respect with non-human others, rather centring the (white, male, Western) human. And lastly, we should remember to guard for the fact that we might not even realise when the systems we build begin to be conscious enough to potentially experience suffering from the tasks we give them and environments we place them in.

## 4. Remembering That Embodiment Matters

AI research has a complex relationship to embodiment. This is most visible with reference to social robots: Sandry (2015) argues that when embodiment is considered, it is too often framed primarily with reference to human communication needs, including speech and physical gestures. In contrast, in present ML work it is vanishly rare to consider algorithms 'embodied'.

It is impossible to parent without recognising that embodiment matters: the early months of parenting (especially for those who give birth and/or mother infants) are shaped by the demands of tending to physical needs. Radical parenting adds to this a recognition of the ways in which social structures shape our interpretation of bodies: through gender, race, and disability, for example.

Bringing this paradigm to AI research may mean more attention to the ways in which race, gender, and other factors (such as international borders) shape the field. At a deeper level, Lewis et al. (2018) argue that, "...AI is formed from not only code, but from materials of the earth. To remove the concept of AI from its materiality is to sever this connection. Forming a relationship to AI, we form a relationship to the mines and the stones. Relations with AI are therefore relations with exploited resources." Crawford and Joler (2018) also provide a striking and beautiful mapping of these embodied relations. Given the enormous impact that the internet, and associated devices, have on the environment, this kind of fundamental rethinking of how we develop new technologies seems overdue.

## 5. Reconsidering Control

The act, or threat, of turning off AGI is often configured as a catalyst for AI beings to become a threat to humans. In the film Ex Machina, there is a dramatized scene where an AI realizes that it is the culmination of such a process; the AI character's reaction is inscrutable, but the audience is left to speculate on whether this teaches it to be equally ruthless in its dealings with humans. Fiction aside, an intelligent being could have a wide range of responses to the realisation that it is being repeatedly deleted and modified by its creators, and not all of them would be traumatised and hostile. But if, as parents, we intend to follow such a path, we should also do our best to ensure that the realisation is not surprising or traumatising. This may also mean we need to watch for whatever it is that causes a discovery to be surprising and unpleasant, and understand when AI systems are beginning to be able to have such reactions to new developments.

We might think of our fears around AGI resisting an off-switch (Hadfield-Menell et al. 2016) as linked to a continuum of control. Algorithms are, in the current paradigm, predominantly mechanisms of control. They are used to manage not only tiny things like the data in computers, but now access to vital resources like healthcare and housing, as well as being deployed in systems of policing and militarisation (Eubanks 2018). AI agents are also, themselves, "imagined as a tool or slave that increases the mana and wealth of 'developers' or 'creators'" (Lewis et al. 2018). Dominant approaches to AI are grounded in, and reinforce, hierarchical relationships. And of course, parenting sometimes pursues similar goals.

The task of understanding one's power as a parent is complex. Conservative approaches tend to position parenting as a process of imposing control, whether through 'negative discipline' or the careful provision of rewards. Radical parenting rejects this balance of punishment and incentives, instead trying to grapple with possessing "a degree of power over the lives of children that we would find inconceivable and unspeakably tyrannical in any other context" (Jordan 2016). Parents must find a way to navigate the reality of having the power to provide or withhold even the most basic needs to their children, as well as finding ways to nurture children's autonomy and, at the same time, protecting them from outside harms and preventing them from harming others.

What might it look like to take this challenge seriously in the case of AI development? Current research programs are comfortable enacting arbitrary degrees of control over and within learning systems, including turning off or deleting agents. Sandry (2018) argues that as our communication with robot others develops, we will need to put more thought into considerations of how humans perceive, and meet the obligations of, 'aliveness'. In one presumably-far-off, improbable future of AGI, this might mean reconsidering the expectation that we will simply switch off beings which we see as disobedient, or even dangerous.

However, even current ML research may benefit from further considerations of how we balance autonomy and control. At some point, as the software and models that we stop, rewrite, re-architect, re-train, or replace become more sophisticated, the use of programmers' current methods may become more morally complex. What is the line between debugging or restarting and retraining a model, and terminating a conscious being in order to replace it with a (hopefully) better one?

Until recently, when programmers have seen their software exhibiting unwanted behaviour, they typically shut down the running program, modify the source code, and start a new copy. In contemporary machine learning contexts, we have gained some new options: one possibility is including the input that triggered the unwanted behaviour or output as an example in a training dataset, and re-training the neural network; another is going back and trying to modify the learning algorithm or model structure so that it is more biased to avoid the unwanted outcome, or more able to learn what not to do. For instance, a neural network that crashed a car (perhaps in a simulation) might be trained on more examples of the road situation that precipitated the crash, or it might be given new heuristics or control theory constraints that are expected to avoid getting close to crashing at all.

Importantly, when children are taught lessons, they continue to be (whether mostly or entirely) the same person. It is much less obvious that this would be true for all of the forms of teaching and education that traditional software, or contemporary neural networks, undergo. Re-training a model with new structures, constraints, or data may involve discarding memories and experiences that the previous model had acquired along the way.

## Conclusion

Parenting theories, and especially radical, queer parenting theories, are an essential addition to the conceptual and analytic frameworks used in AI ethics research. We have found these theories to be rapidly fruitful, shedding new light on old questions and helping us to pose some new ones. We reviewed five of those areas where the parenting lens seems to tell an interesting story: taking great care about how the violence and imperfection of the world is reflected in training data; letting the language and culture of nurture assist both in diversity and progress in the field;

expecting and allowing our AI children to diverge from our forms and values; carefully considering the consequences of imperfect embodiment for the systems we build; and problematizing the degree to which humans assume that they should be able to control AI.

We believe we have only scratched the surface, and that parenting theory (like, say, neuroscience or game theory) is a natural source for experience relevant to important problems in machine learning research, deployment and safety. There are potentials for this approach to be applied poorly: creating a category of under-compensated caring labour within AI research, for example, or even assuming that 'parenting' strategies that seem productive in AI research should be extended to human children. Applied carefully, however, this approach may inspire the creation of new benchmarks and datasets, encourage greater care and responsibility in the design and birth of AI systems, and help to spot more problems before they are shipped to production.

## Acknowledgments

The authors are highly grateful to Elizabeth Przywolnik for excellent assistance and feedback. We thank Brian Christian, Tim Hwang, Kate Crawford, and Eleanor Sandry for passing enthusiasm that encouraged this experimental research effort. We are grateful to the reviewers for their suggestions on further reading and overlooked areas of concern.